\documentclass[aps,prb,preprint,groupedaddress,showpacs,amsmath,amssymb]{revtex4}

\usepackage{graphicx}
\usepackage{dcolumn}
\usepackage{bm}

\begin{document}

\title{Evolution of ground state and upper critical field in R$_{1-x}$Gd$_x$Ni$_2$B$_2$C (R = Lu, Y): Coexistence of superconductivity and spin-glass state}


\author{S. L. Bud'ko}
\author{V. G. Kogan}
\author{H. Hodovanets}
\author{S. Ran}
\author{S. A. Moser}
\author{M. J. Lampe}
\author{P. C. Canfield}
\affiliation{Ames Laboratory, US DOE and Department of Physics and Astronomy, Iowa State University, Ames, Iowa 50011}

\date{\today}

\begin{abstract}

We report  effects of local magnetic moment, Gd$^{3+}$, doping ($x \lesssim 0.3$) on  superconducting and magnetic properties of the closely related  Lu$_{1-x}$Gd$_x$Ni$_2$B$_2$C and Y$_{1-x}$Gd$_x$Ni$_2$B$_2$C series. The superconducting transition temperature decreases and the  heat capacity jump associated with it drops rapidly with Gd-doping; qualitative changes with doping are also observed in the temperature-dependent upper critical field behavior, and a region of coexistence of superconductivity and spin-glass state is delineated on the $x - T$ phase diagram. The evolution of superconducting properties can be understood within Abrikosov-Gor'kov theory of magnetic impurities in superconductors taking into account the paramagnetic effect on upper critical field with additional contributions particular for the family under study.

\end{abstract}


\maketitle

\section{Introduction}
With discoveries of new superconducting materials, classical results on effects of non-magnetic and magnetic impurities in superconductors \cite{and59a,abr60a} are being continuously tested and augmented. For example, the searches for impurity-induced states in superconductors \cite{bal06a} and for superconducting quantum critical points \cite{ram97a,min01a,sha07a} are few such topics. Unfortunately, in some studies of superconductors with magnetic impurities the emphasis is frequently  on just the superconducting properties whereas the state of magnetic subsystem is often neglected.

The rare earth-nickel borocarbides (RNi$_2$B$_2$C, R = rare earth)  present a rare opportunity to study, within the same family, superconductivity, complex, local moment, magnetism, and their coexistence, as well as physics of strongly correlated, heavy fermion materials. \cite{can98a,mul01a,mul02a,bud06a} In this work we concentrate on thermodynamic and magneto-transport properties of LuNi$_2$B$_2$C and YNi$_2$B$_2$C superconductors  with the non-magnetic rare earths (Lu or Y) partially substituted by magnetic moment bearing gadolinium. Pure LuNi$_2$B$_2$C and YNi$_2$B$_2$C have a conveniently high superconducting transition temperatures, $T_c$, and are readily available as well characterized single crystals. The details of the superconducting pairing in these materials are still debated, with exotic scenarios being examined. \cite{wan98a,mak02a,yua03a,yua03b,mak04a}  Since the Gd$^{3+}$ ion has a spherically symmetric, half-filled {\it 4f} shell, and therefore virtually no crystal electric field effects associated with it, using gadolinium as a magnetic rare earth dopant may simplify the problem at hand. Although in resistivity and low field dc magnetic susceptibility the features associated with a magnetic subsystem, if located below $T_c$, are often obscured by strong superconducting signal, it was shown \cite{mov94a,rib03a,bud03a} that in this situation heat capacity measurements can provide a valuable  insight. So far there were several publications, mainly on polycrystalline samples, on physical properties of Y$_{1-x}$Gd$_x$Ni$_2$B$_2$C \cite{mas94a,mas95a,pag98a,lan98a,mic00a,hag00a,drz03a} and Lu$_{1-x}$Gd$_x$Ni$_2$B$_2$C \cite{pag98a,cho96a,rat03a,bud03a} solid solutions. It is noteworthy that although different studies generally agree on the rate of suppression of $T_c$ (on the pure YNi$_2$B$_2$C side) and change of the N\'{e}el temperature, $T_N$, (on the pure GdNi$_2$B$_2$C side) with $x$, separation (absence of coexistence) of the superconducting and antiferromagnetic order in Y$_{1-x}$Gd$_x$Ni$_2$B$_2$C near $x = 0.3$ was alluded to in Ref. \onlinecite{drz03a}, whereas a coexistence of antiferromagnetism and superconductivity at some region of intermediate concentrations was suggested in Refs. \onlinecite{mas95a,hag00a}. Additionally, non-monotonic temperature dependence of the upper critical field, $H_{c2}(T)$, was reported for Lu$_{0.88}$Gd$_{0.12}$Ni$_2$B$_2$C. \cite{rat03a}

A comparative study of the effects of Gd-doping on $T_c$, $H_{c2}(T)$ and  the state of magnetic sublattice in Lu$_{1-x}$Gd$_x$Ni$_2$B$_2$C and Y$_{1-x}$Gd$_x$Ni$_2$B$_2$C, has the potential to clarify the effect of magnetic impurities on the superconducting state in the rare earth-nickel borocarbides.

\section{Experiment}
All samples in this study, Lu$_{1-x}$Gd$_x$Ni$_2$B$_2$C and Y$_{1-x}$Gd$_x$Ni$_2$B$_2$C series, were single crystals, grown using the Ni$_2$B high temperature growth technique \cite{can98a,xum94a,can01a}. {\it As grown} crystals were used for this work. Gd concentrations in both series were evaluated through Curie-Weiss fits of the high-temperature part of magnetic susceptibility, that was measured using a Quantum Design, Magnetic Property Measurement System (MPMS) SQUID magnetometer. For resistance measurements a standard, four probe, ac technique ($f = 16$ Hz, $I = 0.2-2$ mA) with the current flowing in the $ab$ plane, close to $I \| a$, was used. For these measurements platinum wires were attached to the samples using EpoTek H20E silver epoxy and the measurements were performed in a Quantum Design, Physical Property Measurement System (PPMS-14) instrument with ACT and He-3 options. $H_{c2}(T)$ data were obtained from temperature- and magnetic field-dependent resistance measurements. For these measurements $H \| c$ direction of the applied field was kept for all samples. Heat capacity measurements were performed in PPMS-14 instrument with He-3 option utilizing the relaxation technique with fitting of the whole temperature response of the microcalorimeter.

\section{Results and Discussion}
\subsection{Heat Capacity and $x - T$ Phase Diagram}
Since the high temperature paramagnetism in the Lu$_{1-x}$Gd$_x$Ni$_2$B$_2$C and Y$_{1-x}$Gd$_x$Ni$_2$B$_2$C series is associated only with a local moment bearing Gd$^{3+}$ ion, it was expedient to evaluate the real Gd concentration, $x_{CW}$,  by fitting the measured dc susceptibility $\chi_{dc} = M/H$ (between $\sim 150$ K and room temperature) with $\chi_{dc} = x_{CW}C/(T-\Theta)$, where $\Theta$ is the Curie-Weiss temperature, $C = (N_A p_{eff}^2)/3k_B$, $N_A$ is the Avogadro number, $k_B$ is the Boltzmann constant, and $p_{eff}$ is the effective moment (for Gd$^{3+}$ $p_{eff} \approx 7.94 \mu_B$). Fig \ref{F1} shows experimentally evaluated Gd concentration, $x_{CW}$ as a function of the nominal concentration, $x_{nominal}$. For Y$_{1-x}$Gd$_x$Ni$_2$B$_2$C both concentrations are very close to each other ($x_{CW}/x_{nominal} = 1.01(1)$), whereas the difference is fairly large in the case of Lu$_{1-x}$Gd$_x$Ni$_2$B$_2$C ($x_{CW}/x_{nominal} = 1.31(2)$); in both cases the dependence is close to linear in the range of concentrations studied. In the rest of the text, the experimentally determined Gd concentration will be used.

Normalized, zero-field, temperature-dependent  resistivity data, $\rho(T)/\rho_{300K}$, for the  Lu$_{1-x}$Gd$_x$Ni$_2$B$_2$C and Y$_{1-x}$Gd$_x$Ni$_2$B$_2$C series are shown in Fig. \ref{F2}a and  \ref{F2}b. With Gd doping the residual resistivity ratio, RRR =  $\rho_{300K}/\rho_n$, where $\rho_n$ is the normal state resistivity just above the superconducting transition, decreases and the superconducting transition temperature, $T_c$, decreases as well (Fig. \ref{F2}c, inset). The superconducting critical temperature determined from the onset of the resistive superconducting transition for Lu$_{1-x}$Gd$_x$Ni$_2$B$_2$C and Y$_{1-x}$Gd$_x$Ni$_2$B$_2$C  is plotted as a function of Gd concentration in Fig. \ref{F2}c. The $T_c(x)$ dependence is close to linear with a downturn seen in the case of  Y$_{1-x}$Gd$_x$Ni$_2$B$_2$C for the highest presented doping level. This behavior is consistent with Abrikosov-Gor'kov (AG) theory of pairbreaking on magnetic impurities \cite{abr60a}. The rate of $T_c$ suppression is similar for two R$_{1-x}$Gd$_x$Ni$_2$B$_2$C series, being slightly higher for R = Lu. This difference is probably due to the additional contribution of the effect of non-magnetic scattering in  superconductors with anisotropic gap. cite{mar63a,hoh63a,ope97a,kog10a} Indeed, RRR (that can be,  by Matthiessen's rule, roughly taken as a caliper of scattering, with lower RRR corresponding to higher scattering) decreases with Gd doping faster in the case of R = Lu (Fig. \ref{F2}c, inset), that is consistent with larger lattice mismatch (causing stronger scattering) for the Gd/Lu (in comparison to Gd/Y) substitution.   For comparison, the data for $T_c(x)$ evolution in  Lu(Ni$_{1-x}$Co$_x$)$_2$B$_2$C from Ref. \onlinecite{che98a} are included in the same plot. It is noteworthy that the $T_c$ suppression rate is higher for Co-doping to the Ni-site than for Gd-doping  to the Lu(Y) site, even though among local moment rare earth (e.g. excluding Ce and Yb) Gd$^{3+}$  (and Eu$^{2+}$) has the highest de Gennes factor, $(g_J - 1)^2J(J + 1)$, and the strongest $T_c$ suppression rate. \cite{can98a,bud06a}  The reason for such a strong effect of Co-substitution on $T_c$ is at least two-fold: firstly, Co-substitution for Ni is not isoelectronic, it induces changes in the density of states at the Fermi level, therefore causing changes in $T_c$ [\onlinecite{sch94a,bud95a,mat94a}]; secondly, for similar concentrations, $x$, scattering appears to be stronger for Co-substitution (Fig.\ref{F2}c, inset), thus adding to the $T_c$ suppressing rate.

Zero field, temperature-dependent heat capacity, $C_p(T)$, was measured for the  Lu$_{1-x}$Gd$_x$Ni$_2$B$_2$C and Y$_{1-x}$Gd$_x$Ni$_2$B$_2$C series in order to get additional insight into the evolution of the magnetic properties with Gd-doping. The results are presented in Fig. \ref{F3}. For the parent compounds, and several lower Gd concentrations in each series, a jump in $C_p(T)$, at the superconducting transition temperature is clearly seen. This jump broadens with Gd-doping thus the value of $\Delta C_p$ at $T_c$ was evaluated by the isoentropic construct. Fig. \ref{F4} shows the heat capacity jump inferred from the isoentropic construct for the  Lu$_{1-x}$Gd$_x$Ni$_2$B$_2$C and Y$_{1-x}$Gd$_x$Ni$_2$B$_2$C series normalized to the value of the jump for the parent compounds,  LuNi$_2$B$_2$C and YNi$_2$B$_2$C, respectively, plotted as a function of normalized superconducting transition temperature, $T_c/T_{c0}$. As expected, the experimental points lay below the BCS law of corresponding states line, \cite{bar57a,swi59a}  however these points also appear to be below the line obtained within the AG theory of pairbreaking from magnetic impurities \cite{ska64a} as well. Similar behavior of $\Delta C_p/\Delta C_{p0}$ vs  $T_c/T_{c0}$ was observed decades ago for Kondo-impurities (with temperature-dependent pair-breaking)  in superconductors. \cite{mue72a,map76a} In our case the dopant, Gd$^{3+}$, is a good local magnetic moment ion for which hybridization and Kondo-related physics are not expected. There are several possible explanations of such behavior that do not invoke the Kondo effect. Qualitatively similar behavior (approximated by $\Delta C_p \propto T^2$) was observed in  Y$_{1-x}$R$_x$Ni$_2$B$_2$C (R = Gd, Dy, Ho, and Er) \cite{elh00a} and was attributed to a combination of weak-coupling  results of magnetic pairbreaking AG theory with strong coupling corrections. Alternatively, a Hartree-Fock approach by Shiba \cite{shi73a} yields a band of possible  $\Delta C_p/\Delta C_{p0}$ vs  $T_c/T_{c0}$ values that is defined within this approach by the value of the parameter $\gamma$, related to the strength of spin-flip scattering. For $\gamma \to 1$ (weak scattering) the AG results reproduced. The limit of $\gamma \to 0$ describes strong spin-dependent scattering. Our experimental data lay close to this $\gamma \to 0$ limit (Fig. \ref{F4}). Another possible explanation may be a combined effect of magnetic and nonmagnetic scattering \cite{ope04a}  with a notion that the gap parameter in borocarbides is anisotropic. This last possibility is  appealing but requires more theoretical work due to complexity of the theoretical results and a number of independent parameters required for a realistic description. 

Our previous data on the Yb$_{1-x}$Gd$_x$Ni$_2$B$_2$C and Lu$_{1-x}$Gd$_x$Ni$_2$B$_2$C series \cite{bud03a}  provide experimental evidence that for Gd concentration $x \lesssim 0.3$ the long range magnetic order observed in pure GdNi$_2$B$_2$C and the high-Gd end of the series,  evolves into a spin glass (SG).  A broad maximum in heat capacity marked as $T_{max}$ in Fig. \ref{F3}  is associated with a spin glass transition, with $T_{max} \approx 1.5 T_f$, for RKKY spin glasses \cite{myd93a} where $T_f$ is the spin glass freezing temperature. 

The other feature in  temperature-dependent heat capacity data (Fig. \ref{F3})  is a broad minimum. This minimum exists  for all of our $x > 0$ data and is most probably just a crossover between the low temperature magnetism-dominated behavior and high temperature behavior dominated by electron and phonon contributions.

Resistivity and heat capacity data together allow us to construct the $x - T$ phase diagram for the  Lu$_{1-x}$Gd$_x$Ni$_2$B$_2$C and Y$_{1-x}$Gd$_x$Ni$_2$B$_2$C series (Fig. \ref{F5}).  As mentioned above, there is a slight difference in $T_c$ variation with $x$ between R = Lu and R = Y. The other salient temperature, $T_{max}$,  has very similar $x$ - dependence in both cases. It has to be mentioned that probing magnetic signatures at temperatures below superconducting transition often is not a simple task. In electric/thermoelectric  and low field magnetic susceptibility measurements the superconducting signal dominates. Magnetic field needed to suppress superconductivity might be large enough to alter fragile, low temperature, magnetic state (as it happens e.g. in materials with field-induced quantum critical point \cite{ste01a}), or at a minimum, shift the phase line. Zero-field heat capacity measurements clearly reveal (complex) long range magnetic order below $T_c$. \cite{cho95a,rib03a} In the case of spin-glass transition heat capacity does not have clear anomaly at the freezing temperature, $T_f$, instead a broad maximum is detected at $\approx 1.5 T_f$. \cite{myd93a} Having this in mind, we can approximately outline (by the dotted-dashed line in Fig. \ref{F5}) the boundary of the spin-glass phase. Since, at least in zero-field resistivity, that was measured in this work down to the temperatures below the SG line for several Gd-concentrations, no reentrance behavior is observed, superconductivity coexists with the SG state at low temperatures. For slightly higher Gd concentrations, after superconductivity is just suppressed, (as it was mentioned for  Lu$_{1-x}$Gd$_x$Ni$_2$B$_2$C \cite{bud03a}) spin-glass related behavior is observed both in heat capacity and magnetic susceptibility. On further Gd-doping, a long range magnetic order is established.

\subsection{Upper Critical Field}

The upper critical field was measured resistively, combining magnetic field-dependent data taken at constant temperature and temperature-dependent data taken in fixed magnetic field. Examples of such data for  Lu$_{0.81}$Gd$_{0.19}$Ni$_2$B$_2$C ($H \| c$) are shown in Figs. \ref{F6} (a) and (b). Re-entrant $R(T)$ curves  for a few, relatively high, values of magnetic field (Fig. \ref{F6}(b)) are noteworthy. Results obtained from both  data sets  are consistent, the resulting $H_{c2}(T)$ curves for two different criteria are shown in Fig. \ref{F6}(c). The aforementioned re-entrant $R(T)$ curves are the results of the horizontal ($H = constant$) cuts through the shallow maximum in the $H_{c2}(T)$.

The $H_{c2}(T)$ data for the  Lu$_{1-x}$Gd$_x$Ni$_2$B$_2$C and Y$_{1-x}$Gd$_x$Ni$_2$B$_2$C series are presented in Fig. \ref{F7}. The evolution of the upper critical field behavior with Gd-doping is similar for both series: the behavior changes from monotonic with temperature for the parent and lightly-doped compounds to the behavior with shallow maximum for higher Gd concentrations. This evolution is seen better yet when plotted in normalized coordinates (Fig. \ref{F8}). Qualitatively similar evolution of $H_{c2}(T)$ was theoretically described (in dirty limit) by taking into account paramagnetic effect. {\cite{mak64a,sai69a} The use of dirty limit for this materials is consistent with previous studies. \cite{che98a}. The quantitative description of $H_{c2}(T)$ within a paramagnetic effect approach requires detailed knowledge of the paramagnetic contribution to susceptibility below $T_c$, which is a tedious task. On careful examination of Fig. \ref{F8}(b) we can see that a noticeable broad maximum in $H_{c2}(T)$ is observed for $x = 0.14$ and $x = 0.21$, however this maximum practically disappears for the next concentration, $x = 0.26$, for which $H_{c2}(T)$ is monotonic with a tendency to saturation below $T/T_c \approx 0.5$. For this concentration (at zero field) the $T_c$ value is close to $T_f$, the SG freezing temperature (Fig. \ref{F5}). For spin glasses the paramagnetic component of susceptibility decreases below $T_f$ [\onlinecite{myd93a}] so that $H_{c2}$ suppression is expected to be weaker, in agreement with our observation. Similar arguments were used in Ref. \onlinecite{buz86a} for interpretation of $H_{c2}(T)$ data below the N\'eel temperature.

Fig. \ref{F9} presents the slope of the $H_{c2}(T)$ in the limit of $H \to 0$ as a function of $T_c$ in zero field for the  Lu$_{1-x}$Gd$_x$Ni$_2$B$_2$C and Y$_{1-x}$Gd$_x$Ni$_2$B$_2$C series. The observed behavior can be roughly approximated as $dH_{c2}/dT \propto T_c$. It in noteworthy that for several recently studied superconductors the behavior is qualitatively different: for Ce$_{1-x}$La$_x$CoIn$_5$  $dH_{c2}/dT$ is approximately constant for $H \|  c$ and has a factor of two larger absolute value with a slight positive slope for $H \| a$; \cite{pet02a} for neutron-irradiated MgB$_2$ $dH_{c2}/dT$ is approximately independent of $T_c$, \cite{wil06a} whereas for carbon-doped MgB$_2$  $|dH_{c2}/dT|$ rapidly increases with decrease of $T_c$  [\onlinecite{wil04a}] (opposite to what is observed here); for Co-doped LuNi$_2$B$_2$C the derivative decreases in the absolute value only by $\approx 20\%$ when $T_c$ decreases approximately by half. \cite{che98a}

It is worth mentioning that $dH_{c2}/dT \propto T_c$ is predicted for isotropic $s$-wave materials in the clean limit. As discussed above, such description appear not to be pertinent to the borocarbides.  On the other hand, such proportionality is a property of the AG gapless state \cite{abr60a,kog10a,kog09a}  and is present (at least approximately) in the data from elemental La doped with Gd [\onlinecite{cha72a}] (the data in the publication need to be reanalyzed to extract the derivatives). Recently, similar behavior was observed in 1111 family of Fe-As superconductors and was attributed to pair-breaking in anisotropic superconductors. \cite{kog09a} 

Following Ref. \onlinecite{kog09a}, $|d (dH_{c2}/dT)/dT_c| \propto \pi \phi_0 k_B^2/\hbar^2 v^2$, where $\phi_0$ is the flux quantum, $k_B$ is the Boltzmann constant and $v$ is the Fermi velocity. Since  $|d (dH_{c2}/dT)/dT_c| \approx 0.25$ kOe/K$^2$ (Fig. \ref{F9}), the order of magnitude estimate gives $v \sim 3~\times~ 10^7$ cm/s. This estimate is consistent with the values used to describe superconductivity in parent LuNi$_2$B$_2$C and YNi$_2$B$_2$C. \cite{kog97a,shu98a}

\section{Summary}

Gd-doping of  LuNi$_2$B$_2$C and YNi$_2$B$_2$C results in  $T_c$ suppression, consistent with AG magnetic pairbreaking with possible additional contribution from non-magnetic scattering in materials with anisotropic gaps. For both series $T_c$  is suppressed to zero by 30-35\% Gd substitution. The $x - T$ phase diagram reveals a region of co-existence between superconductivity and a spin-glass state arising from the Gd-magnetism. The evolution of the temperature-dependent $H_{c2}$ with Gd-doping can be understood by taking into account the paramagnetic effect and, for the superconducting sample with highest Gd-concentration in this study,  Y$_{0.74}$Gd$_{0.26}$Ni$_2$B$_2$C, by considering temperature dependence of paramagnetic susceptibility below the SG freezing temperature. The $H_{c2}$ derivatives in the limit of $H \to 0$ are approximately linear with zero-field superconducting transition temperatures, in agreement with the behavior expected for AG pairbreaking.

All in all, the  Lu$_{1-x}$Gd$_x$Ni$_2$B$_2$C and Y$_{1-x}$Gd$_x$Ni$_2$B$_2$C series  present viable systems for studies of magnetic pairbreaking in anisotropic superconductors and interplay of superconductivity and spin glass state. 

\begin{acknowledgments}
Work at the Ames Laboratory was supported by the US Department of Energy - Basic Energy Sciences under Contract No. DE-AC02-07CH11358. This manuscript was finalized during the Ames floods of 2010, the second "hundred year floods" in a 15 year time span.
\end{acknowledgments}

\clearpage

\begin{figure}
\begin{center}
\includegraphics[angle=0,width=120mm]{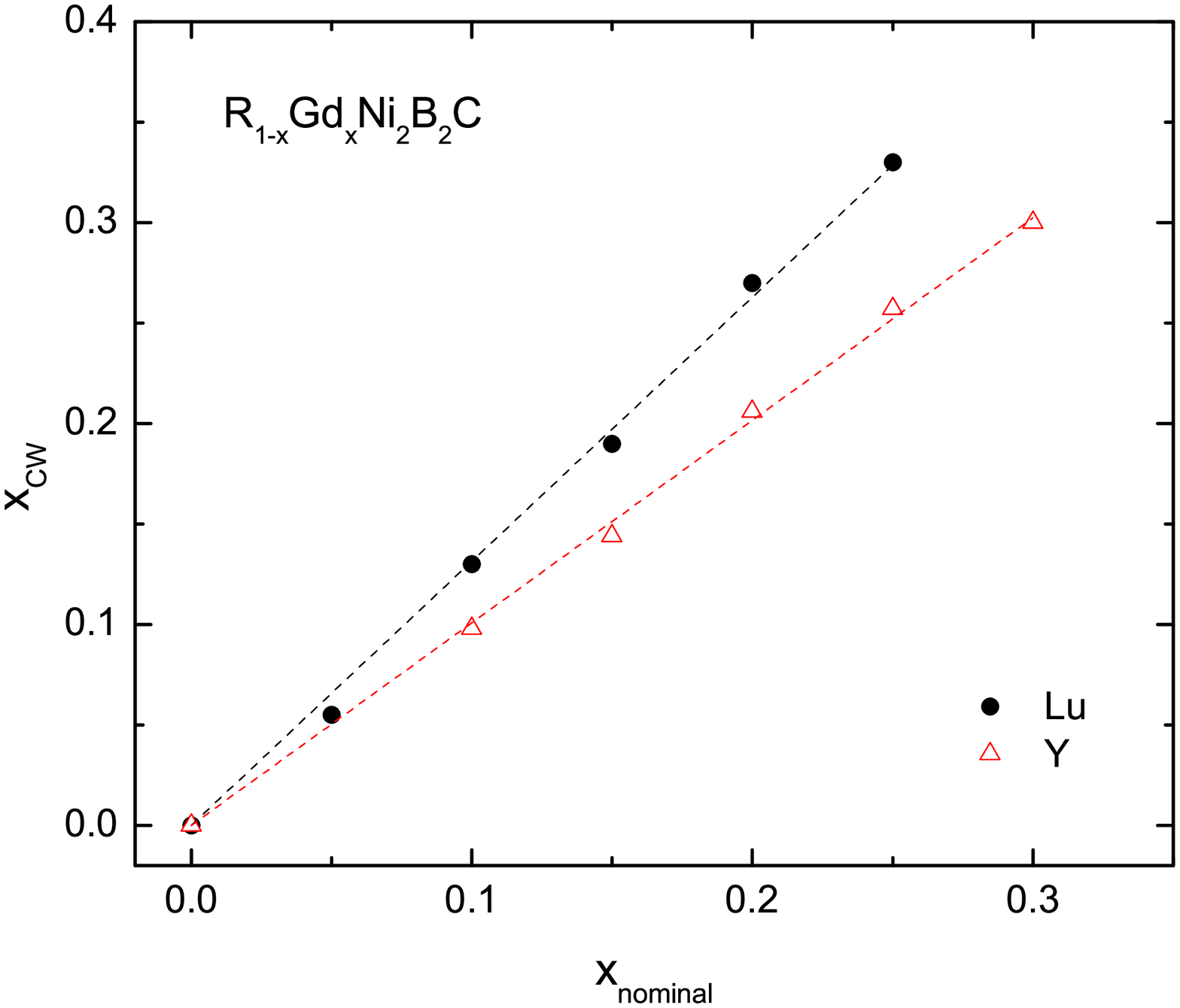}
\end{center}
\caption{(Color online) Gd-concentration evaluated from a Curie-Weiss fit of the high temperature susceptibility vs nominal Gd-concentration in R$_{1-x}$Gd$_x$Ni$_2$B$_2$C, R = Lu, Y. Dashed lines are linear fits with intercept fixed to zero. (see text for details)}\label{F1}
\end{figure}

\clearpage

\begin{figure}
\begin{center}
\includegraphics[angle=0,width=80mm]{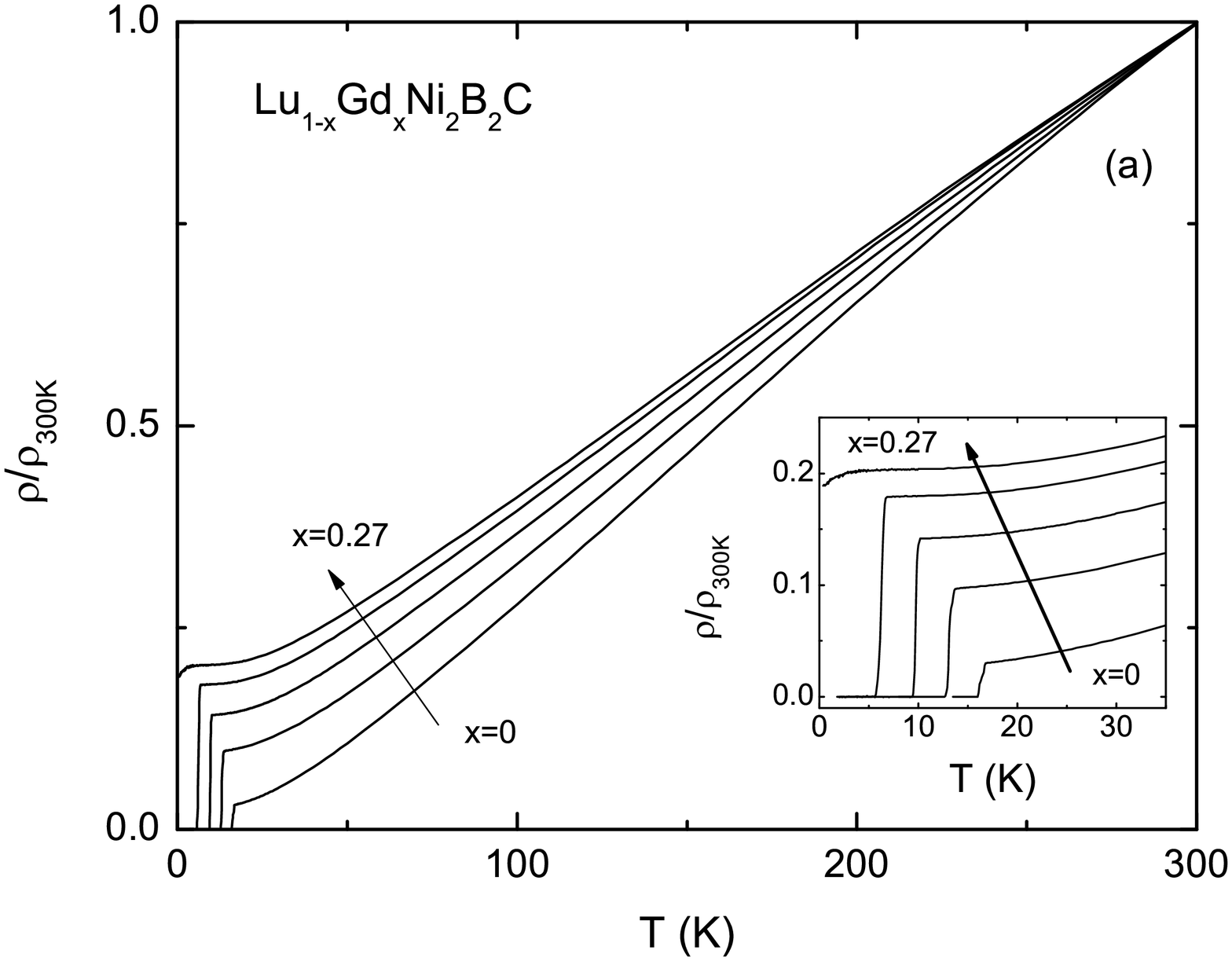}
\includegraphics[angle=0,width=80mm]{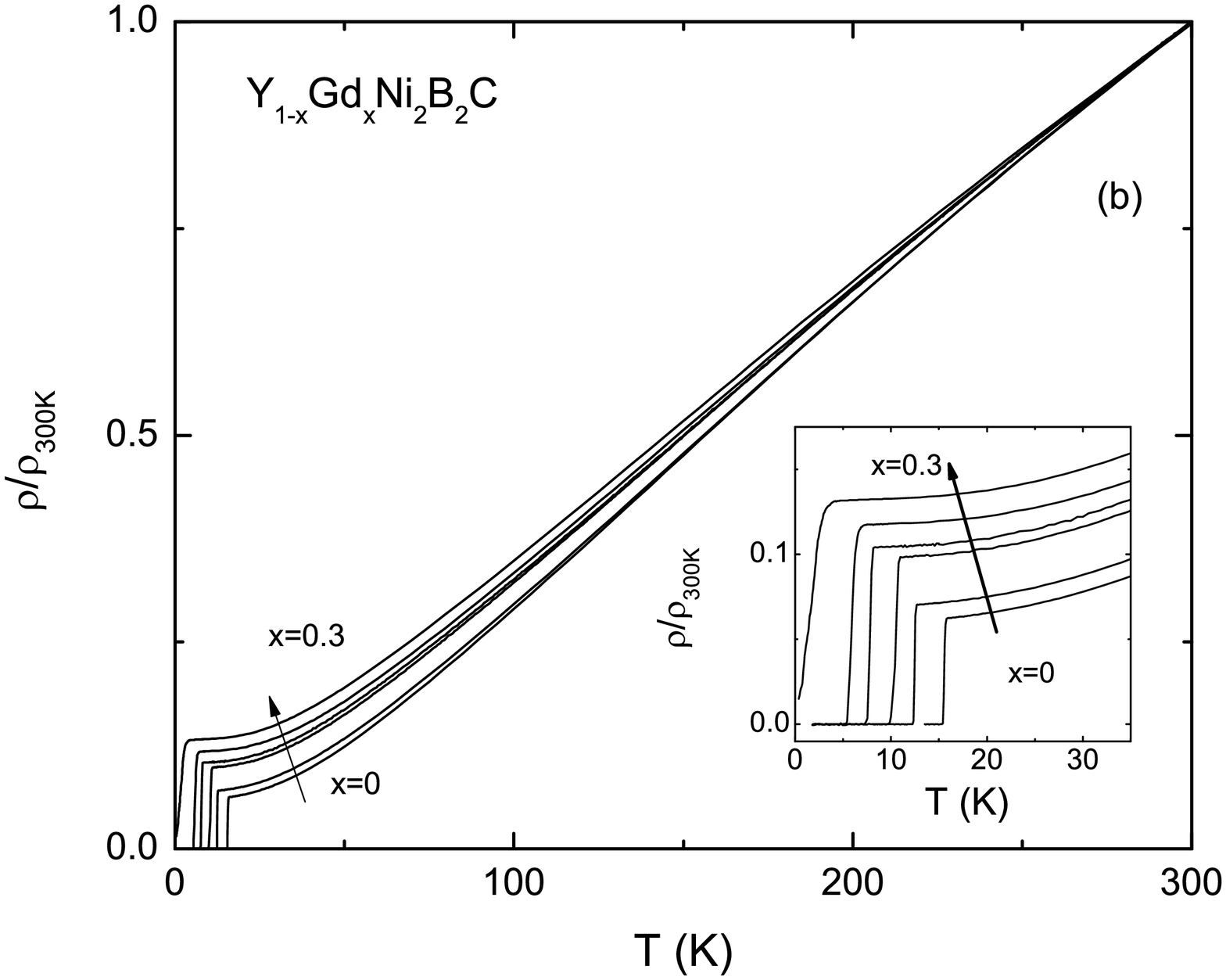}
\includegraphics[angle=0,width=80mm]{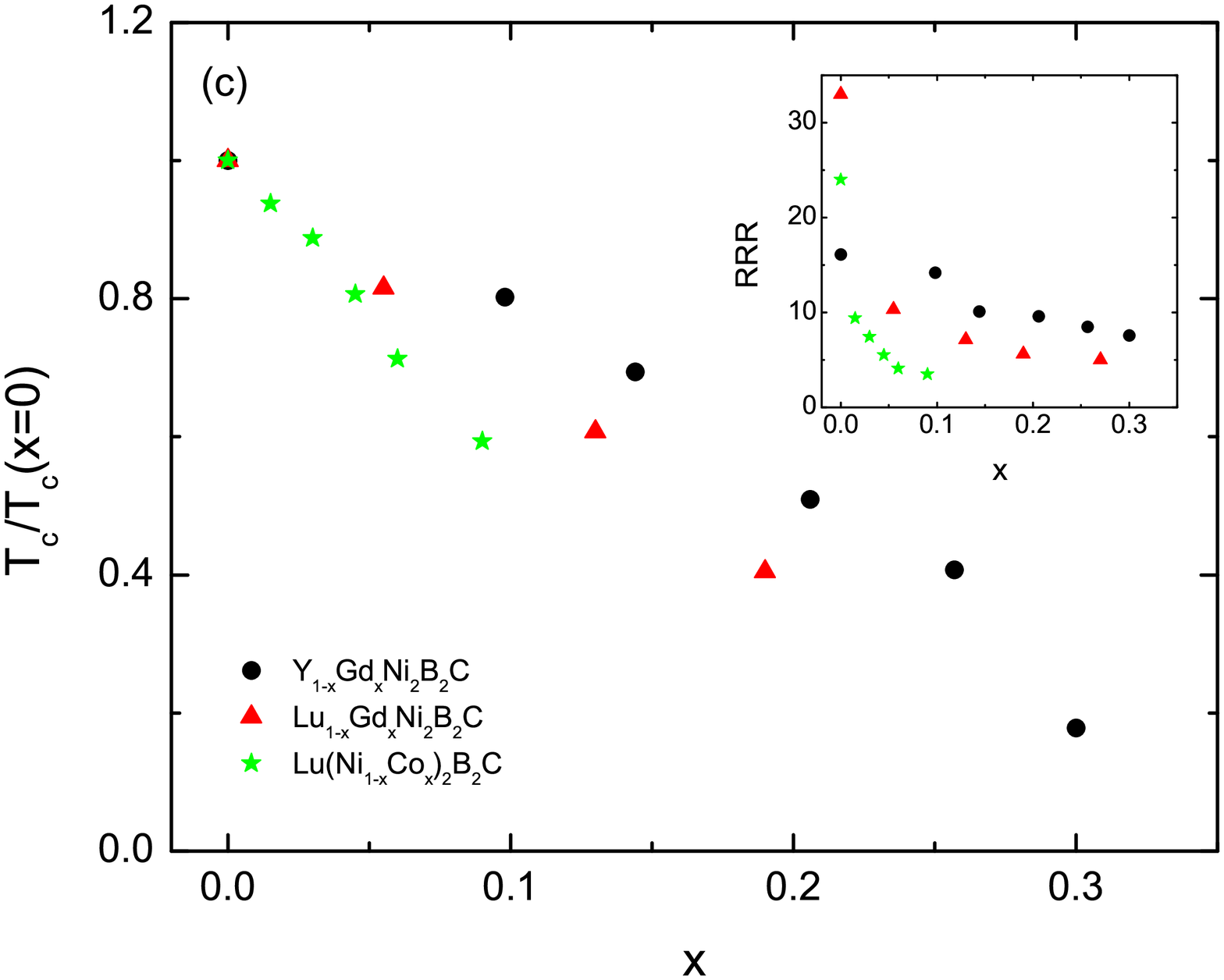}
\end{center}
\caption{(Color online) Normalized resistivity, $\rho/\rho_{300K}$ for (a) Lu$_{1-x}$Gd$_x$Ni$_2$B$_2$C ($x$ = 0, 0.055, 0.13, 0.19, 0.27), and (b) Y$_{1-x}$Gd$_x$Ni$_2$B$_2$C ($x$ = 0, 0.10, 0.14, 0.21, 0.26, 0.30). Arrows show the direction of increasing $x$, insets:  low temperature part of the data. Panel (c):  Normalized (to the values for the parent compounds)  $T_c$ as a function of Gd concentration $x$ for  R$_{1-x}$Gd$_x$Ni$_2$B$_2$C, R = Lu, Y; data for Lu(Ni$_{1-x}$Co$_x$)$_2$B$_2$C from Ref. \onlinecite{che98a} are included for comparison. Inset: RRR vs $x$ for the same three series.}\label{F2}
\end{figure}

\clearpage

\begin{figure}
\begin{center}
\includegraphics[angle=0,width=120mm]{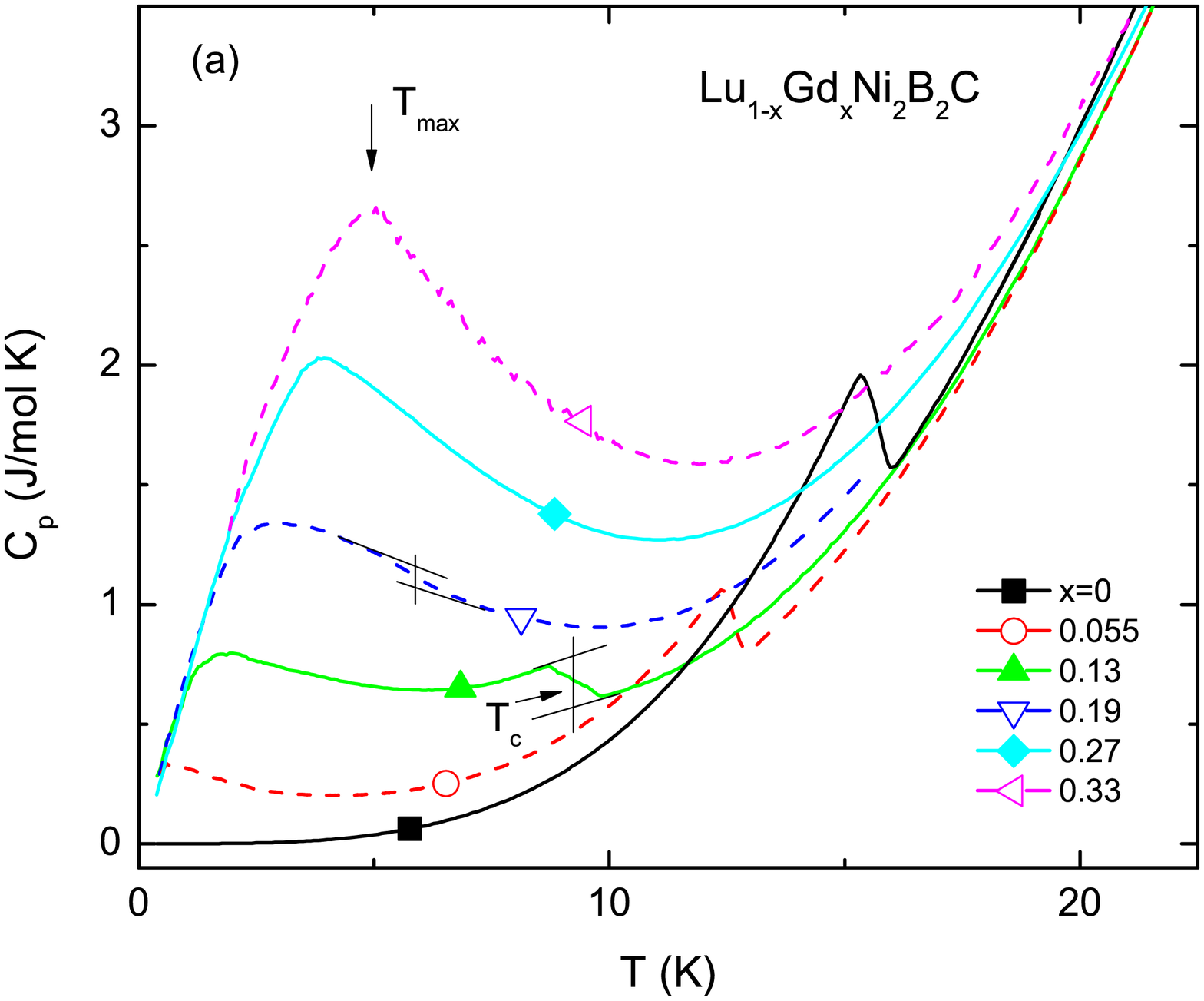}
\includegraphics[angle=0,width=120mm]{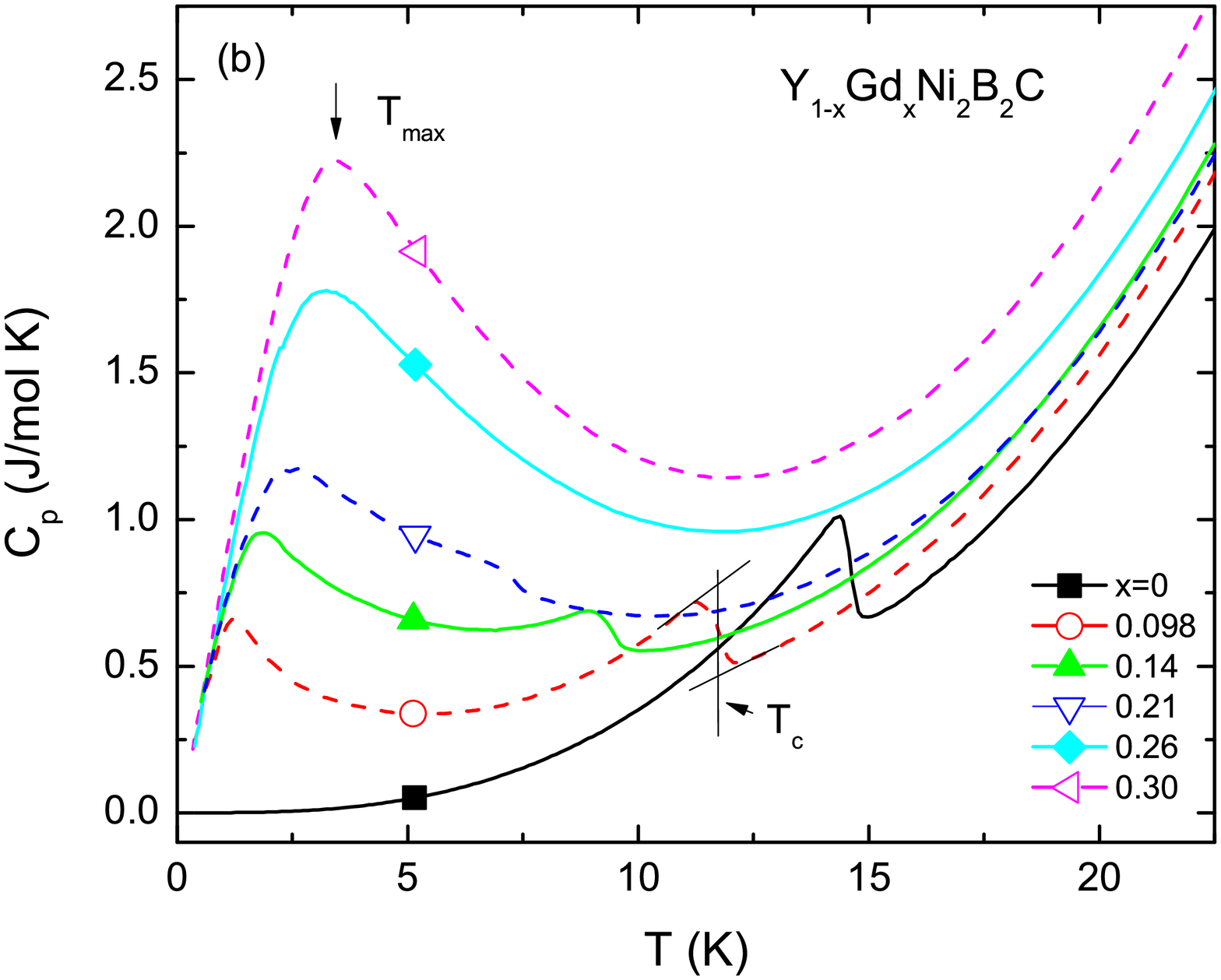}
\end{center}
\caption{(Color online) Temperature dependent heat capacity for (a) Lu$_{1-x}$Gd$_x$Ni$_2$B$_2$C ($x$ = 0, 0.055, 0.13, 0.19, 0.27, 0.33), and (b) Y$_{1-x}$Gd$_x$Ni$_2$B$_2$C ($x$ = 0, 0.10, 0.14, 0.21, 0.26, 0.30). Arrows show examples of how $T_{max}$ and $T_c$ are determined.}\label{F3}
\end{figure}

\clearpage

\begin{figure}
\begin{center}
\includegraphics[angle=0,width=120mm]{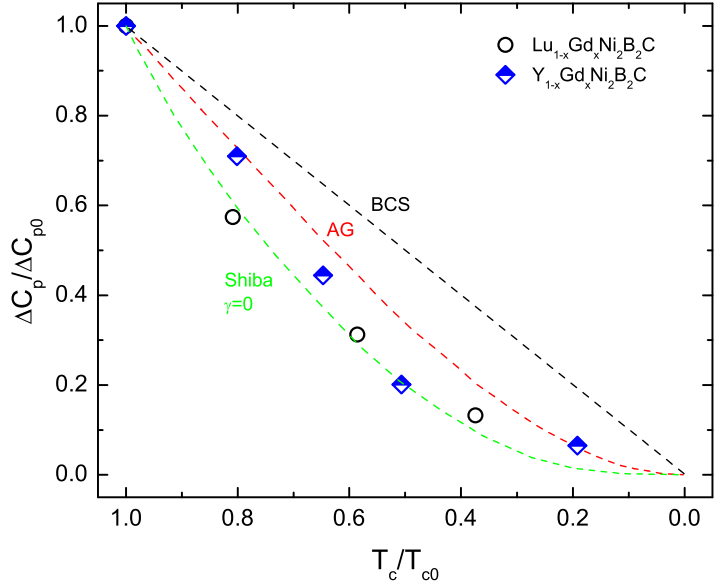}
\end{center}
\caption{(Color online) Normalized jump in heat capacity at $T_c$ vs normalized $T_c$ for the  Lu$_{1-x}$Gd$_x$Ni$_2$B$_2$C and Y$_{1-x}$Gd$_x$Ni$_2$B$_2$C series. Dashed lines correspond to BCS law of corresponding states, Abrikosov-Gor'kov magnetic scattering, and $\gamma = 0$ (strong spin-dependent scattering) limit of Shiba's theory.\cite{shi73a} See text for more details.}\label{F4}
\end{figure}

\clearpage

\begin{figure}
\begin{center}
\includegraphics[angle=0,width=120mm]{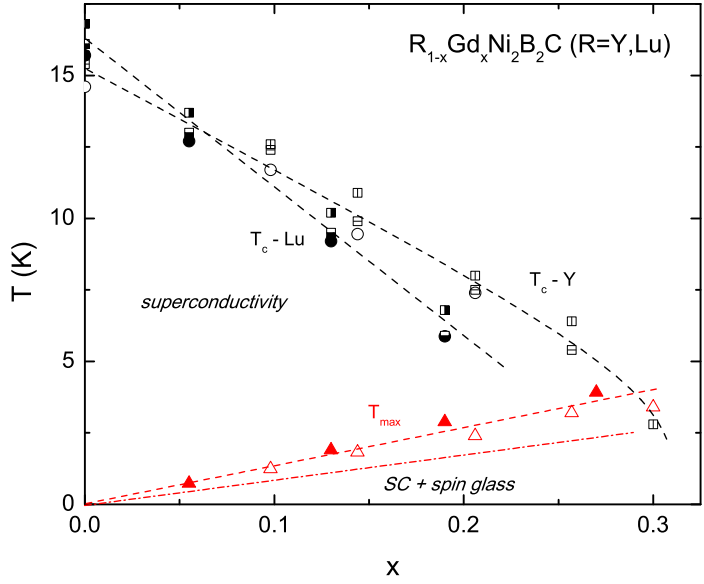}
\end{center}
\caption{(Color online) $T - x$ phase diagram for the  Lu$_{1-x}$Gd$_x$Ni$_2$B$_2$C (filled and partially filled symbols) and Y$_{1-x}$Gd$_x$Ni$_2$B$_2$C (open symbols) series. Symbols: squares - $T_c$ from onset and offset of the resistive transitions; circles - $T_c$ from heat capacity; triangles - $T_{max}$  from heat capacity. Dashed lines are guides for the eye. Dotted-dashed line approximates spin-glass phase (crossover) line.}\label{F5}
\end{figure}

\clearpage

\begin{figure}
\begin{center}
\includegraphics[angle=0,width=80mm]{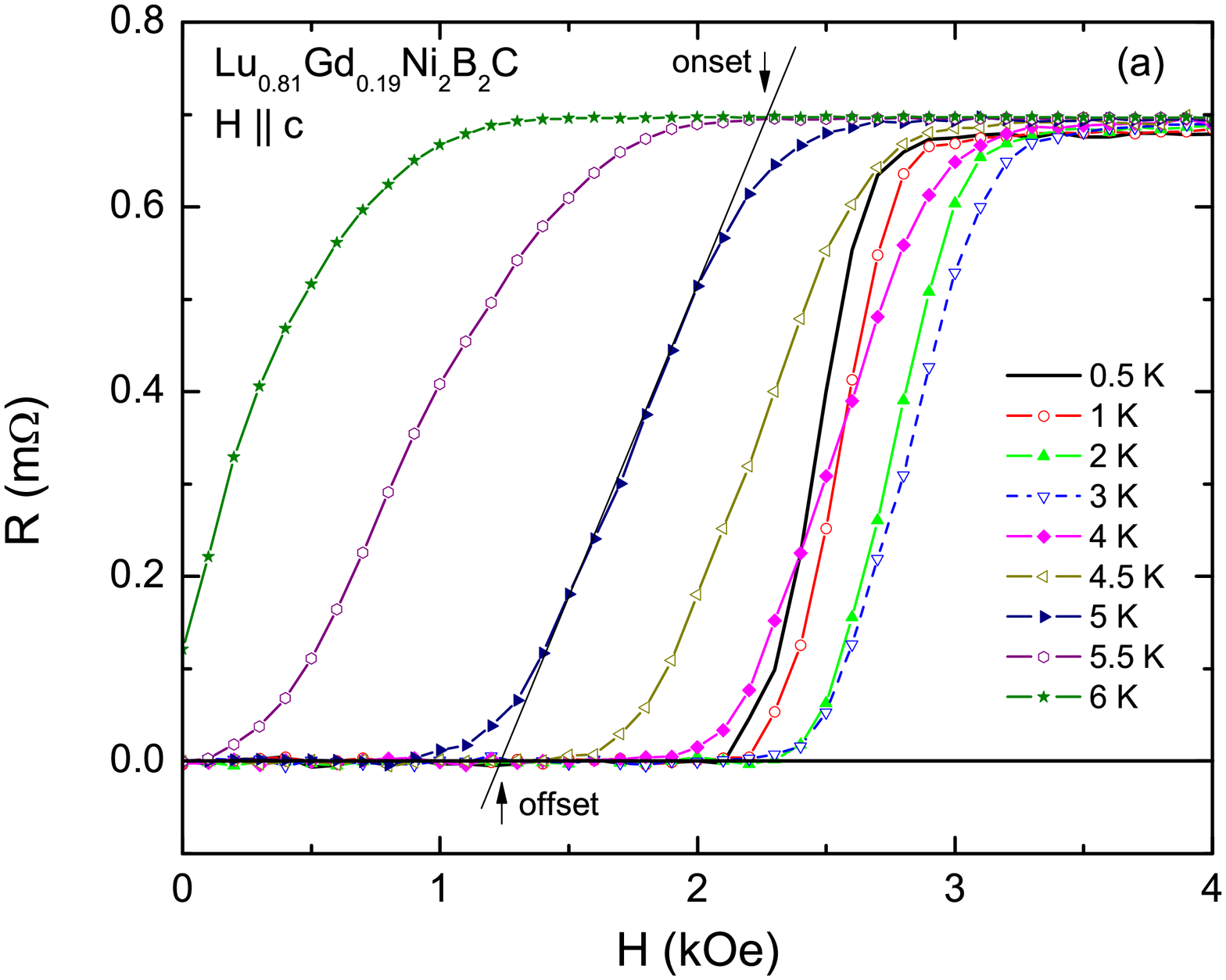}
\includegraphics[angle=0,width=80mm]{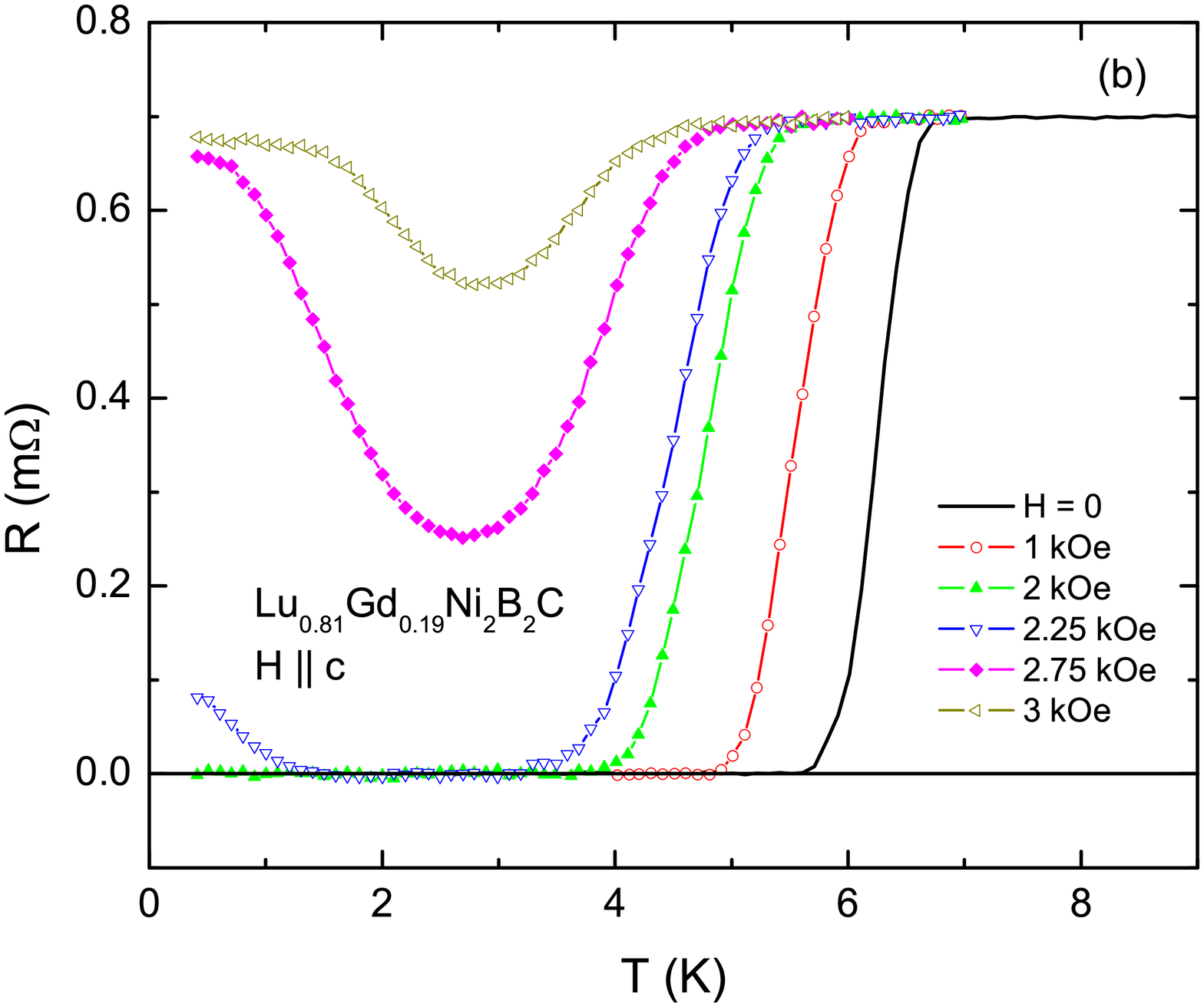}
\includegraphics[angle=0,width=80mm]{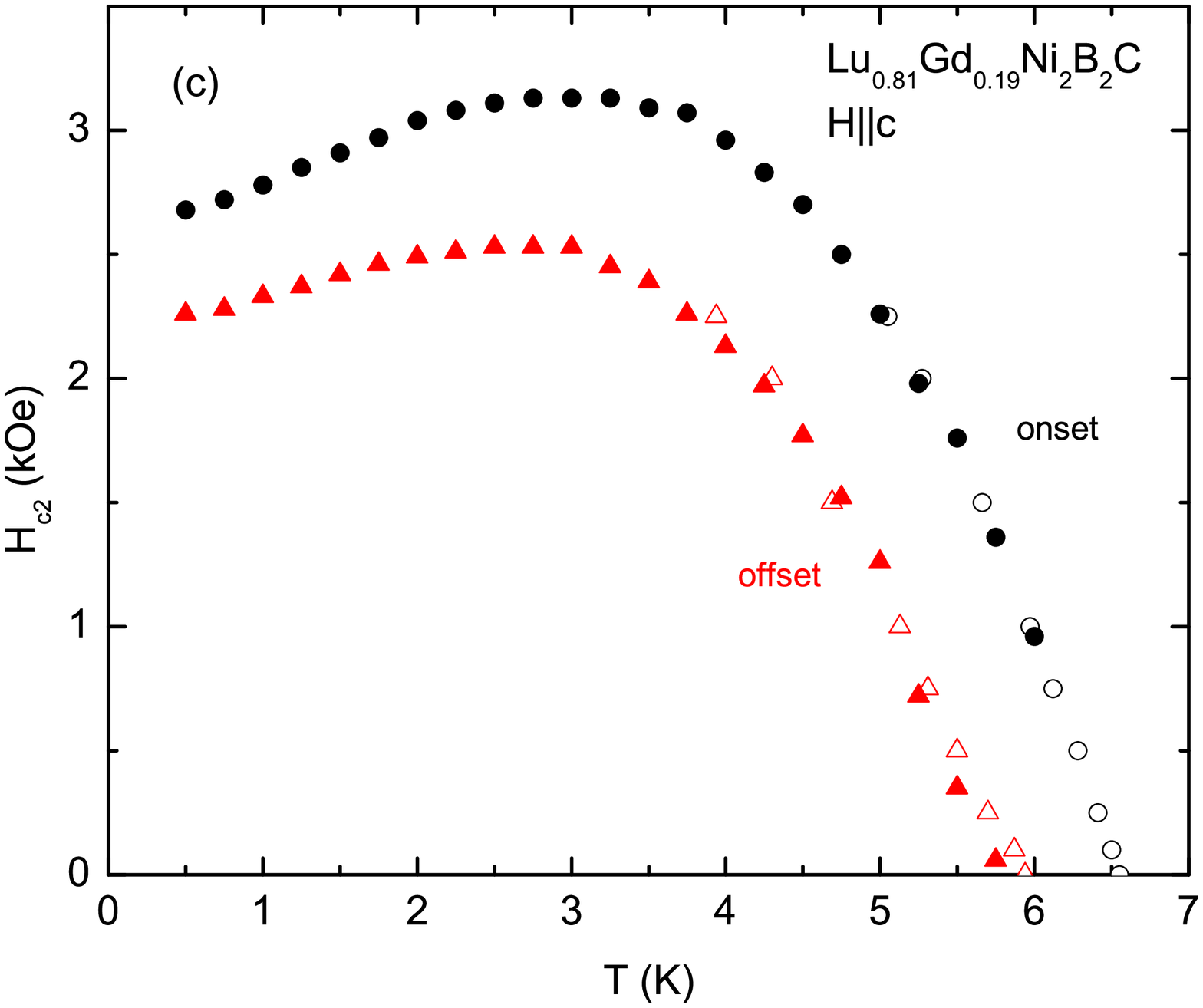}
\end{center}
\caption{(Color online) (a) Examples of magnetic field-dependent resistance of  Lu$_{0.81}$Gd$_{0.19}$Ni$_2$B$_2$C single crystal measured at several constant temperatures for $H \| c$. Onset and offset criteria of superconducting transition are illustrated. (b) Examples of temperature-dependent resistance of the same sample measured in different applied magnetic fields. (c) Temperature-dependent upper critical field of  Lu$_{0.81}$Gd$_{0.19}$Ni$_2$B$_2$C for $H \| c$. Circles - onset, triangles - offset, open symbols are from $R(T)|_H$ scans, filled symbols are from $R(H)|_T$ scans.}\label{F6}
\end{figure}

\clearpage

\begin{figure}
\begin{center}
\includegraphics[angle=0,width=120mm]{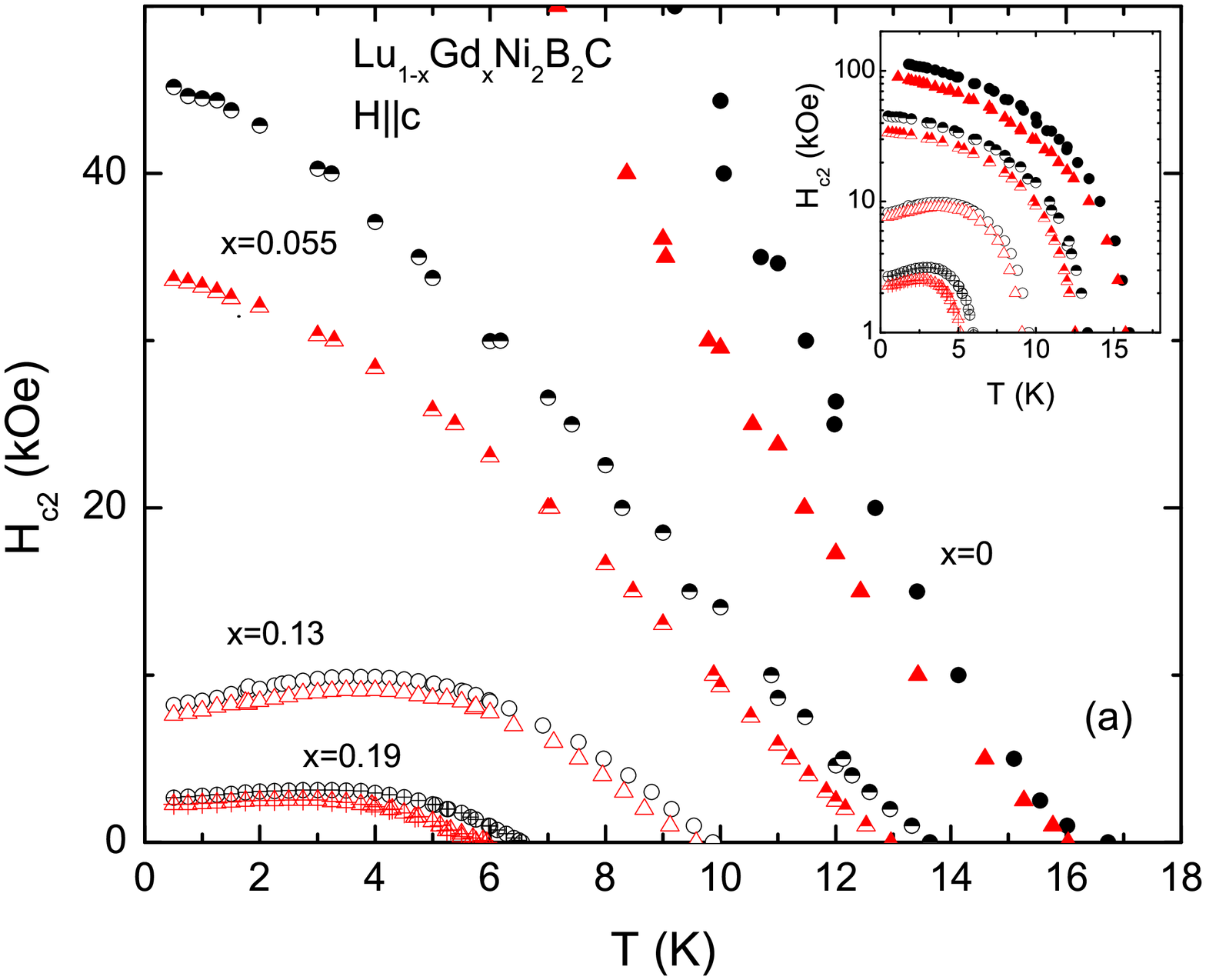}
\includegraphics[angle=0,width=120mm]{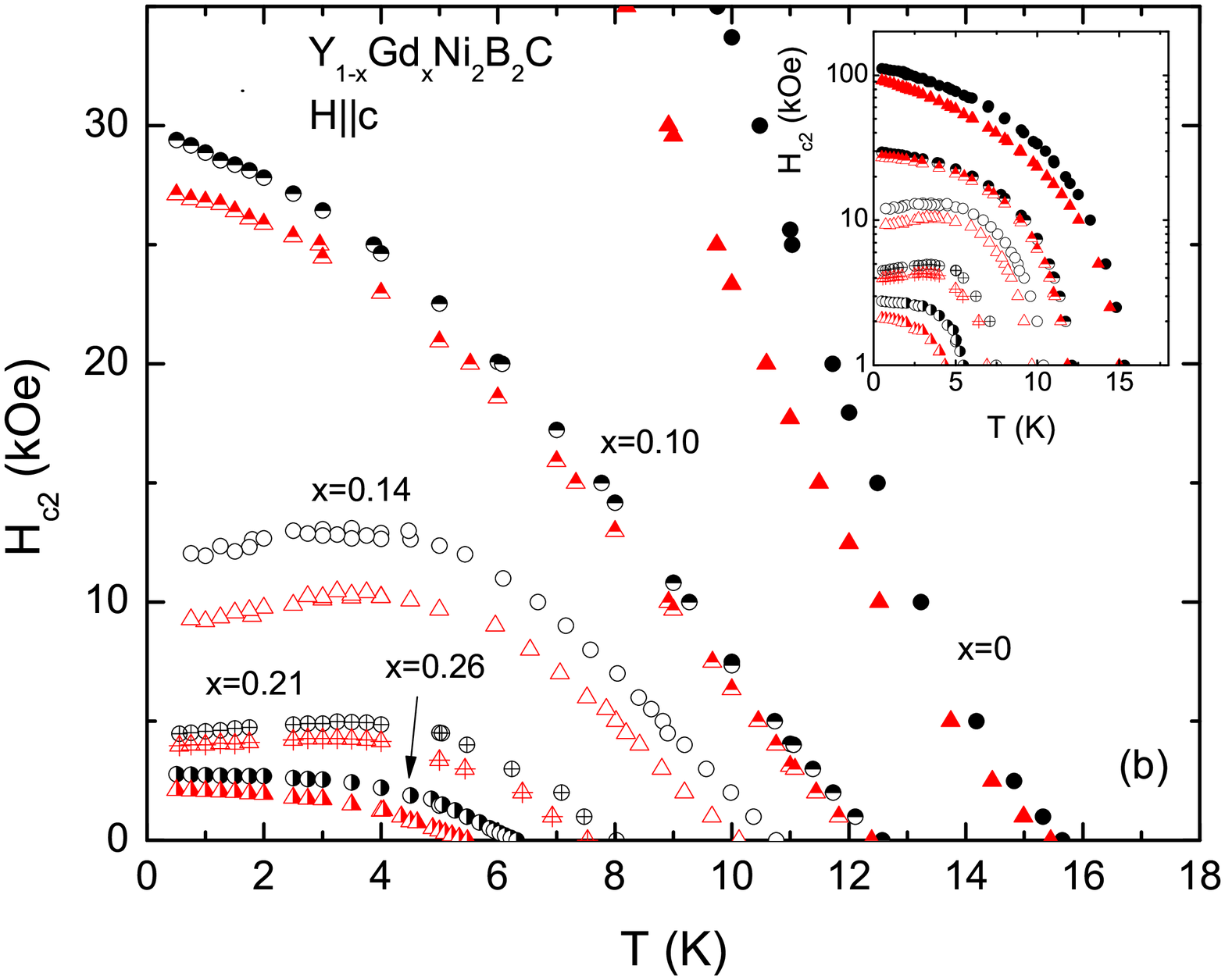}
\end{center}
\caption{(Color online) Temperature dependent upper critical field ($H \| c$) for (a) Lu$_{1-x}$Gd$_x$Ni$_2$B$_2$C ($x$ = 0, 0.055, 0.13, 0.19), and (b) Y$_{1-x}$Gd$_x$Ni$_2$B$_2$C ($x$ = 0, 0.10, 0.14, 0.21, 0.26). Circles and triangles correspond to the onset and offset criteria respectively. Insets show the same data on a semi-log scale.}\label{F7}
\end{figure}

\clearpage

\begin{figure}
\begin{center}
\includegraphics[angle=0,width=120mm]{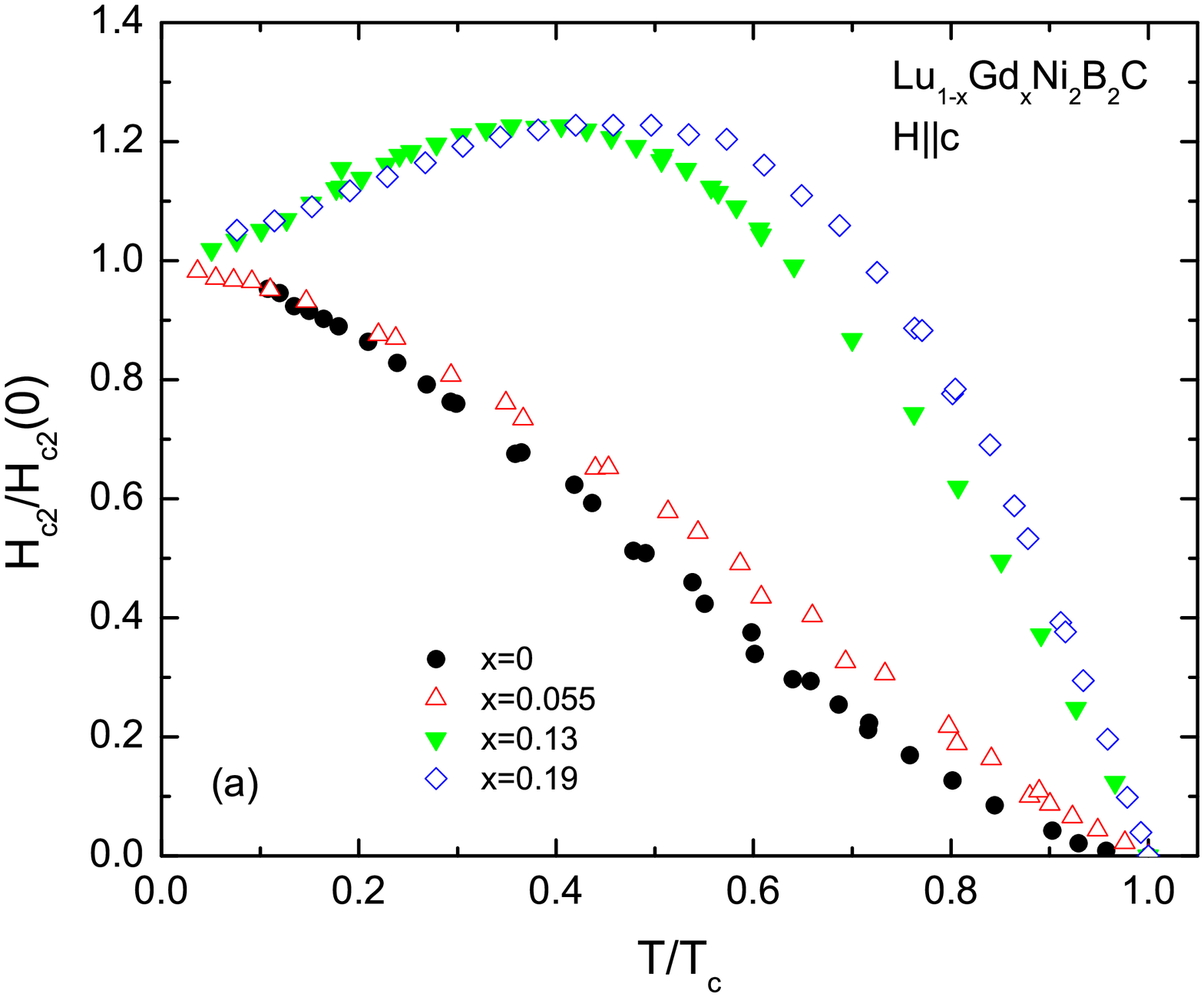}
\includegraphics[angle=0,width=120mm]{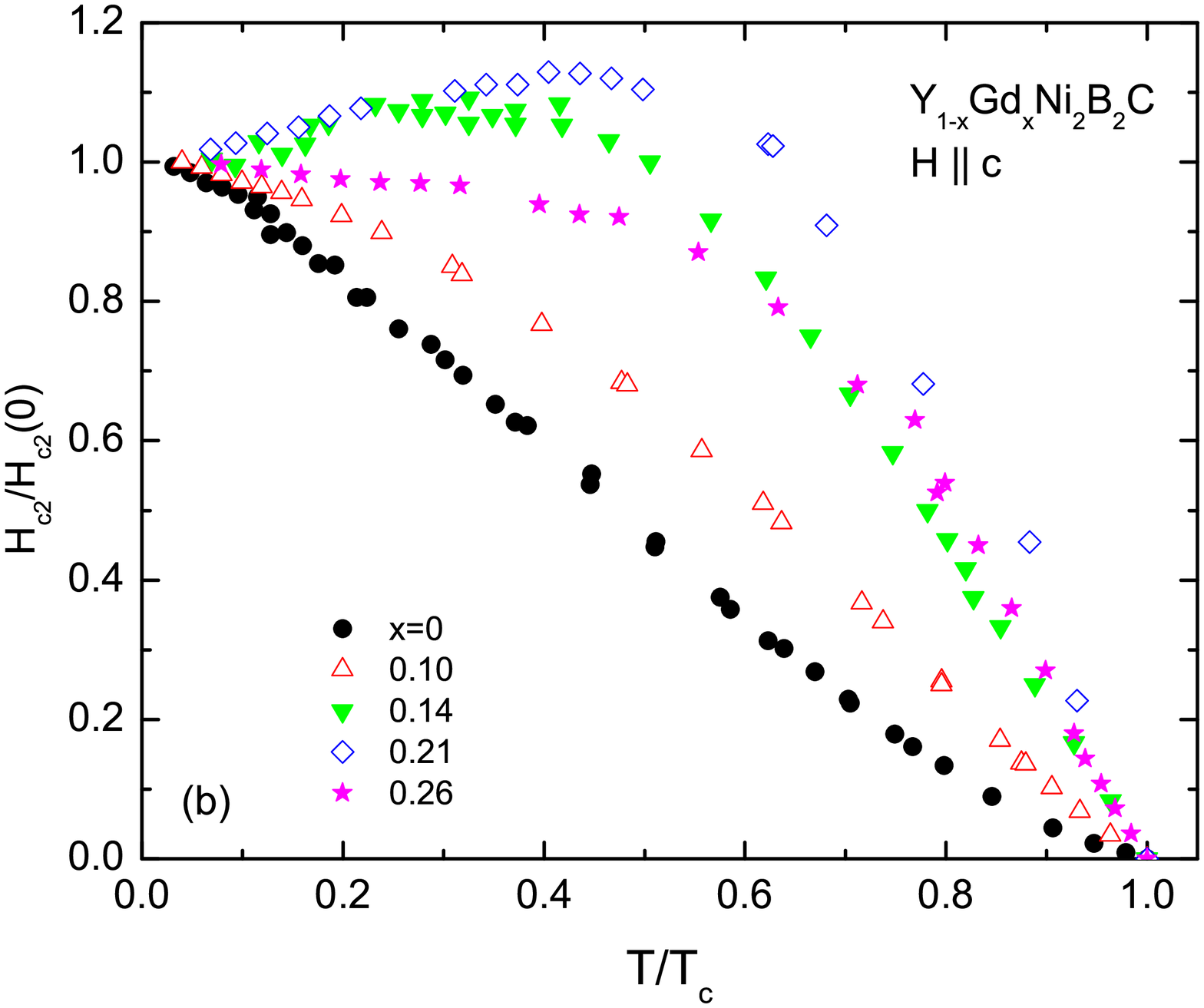}
\end{center}
\caption{(Color online) Normalized to $H_{c2}(T=0)$ temperature dependent upper critical field for (a) Lu$_{1-x}$Gd$_x$Ni$_2$B$_2$C ($x$ = 0, 0.055, 0.13, 0.19), and (b) Y$_{1-x}$Gd$_x$Ni$_2$B$_2$C ($x$ = 0, 0.10, 0.14, 0.21, 0.26) as a function of normalized to $T_c(H=0)$ temperature. Data for $H \| c$ obtained using onset criteria are shown.}\label{F8}
\end{figure}

\clearpage

\begin{figure}
\begin{center}
\includegraphics[angle=0,width=120mm]{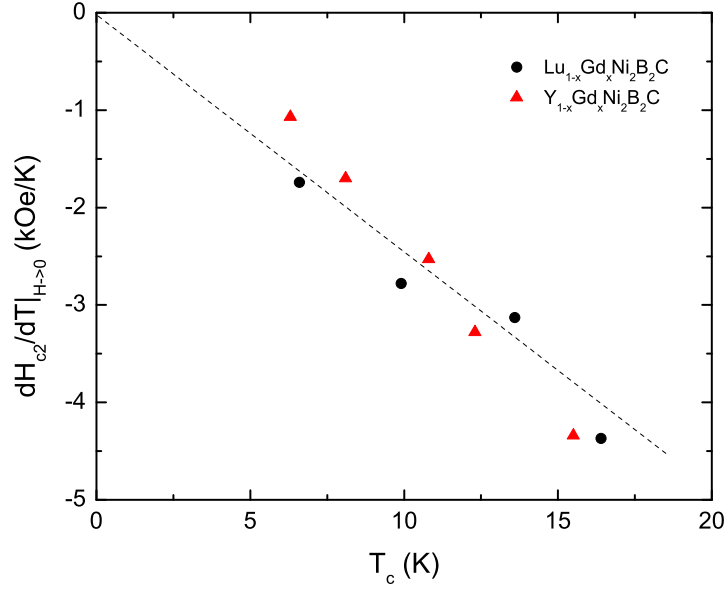}
\end{center}
\caption{(Color online) $dH_{c2}/dT$ in the $H \to 0$ limit as a function of $T_c$ in zero field for the  Lu$_{1-x}$Gd$_x$Ni$_2$B$_2$C (circles) and Y$_{1-x}$Gd$_x$Ni$_2$B$_2$C (triangles) series. Dashed line is a guide to the eye. }\label{F9}
\end{figure}

\end{document}